\documentclass[aps,twocolumn,superscriptaddress]{revtex4-2}

\usepackage{amsmath}    
\usepackage{graphicx}   
\usepackage{verbatim}   
\usepackage{color}      
\usepackage{hyperref}   

\begin{document}

\title{Ultrafast electron dynamics in Au/Fe/MgO(001) analyzed by Au- and Fe-selective pumping in time-resolved two-photon photoemission spectroscopy: Separation of excitations in adjacent metallic layers}

\author{Y. Beyazit}
\affiliation{Faculty of Physics and Center for Nanointegration
(CENIDE), University of Duisburg-Essen, Lotharstr.~1, 47057
Duisburg, Germany}

\author{F. K\"{u}hne}
\affiliation{Faculty of Physics and Center for Nanointegration
(CENIDE), University of Duisburg-Essen, Lotharstr.~1, 47057
Duisburg, Germany}

\author {D. Diesing}
\affiliation{Faculty of Chemistry, University of Duisburg-Essen, Universit\"{a}tsstr. 5, 45711 Essen, Germany}

\author{P. Zhou}
\affiliation{Faculty of Physics and Center for Nanointegration
(CENIDE), University of Duisburg-Essen, Lotharstr.~1, 47057
Duisburg, Germany}

\author{J. Jayabalan}
\affiliation{Faculty of Physics and Center for Nanointegration
(CENIDE), University of Duisburg-Essen, Lotharstr.~1, 47057
Duisburg, Germany}

\author{B. Sothmann}
\affiliation{Faculty of Physics and Center for Nanointegration
(CENIDE), University of Duisburg-Essen, Lotharstr.~1, 47057
Duisburg, Germany}

\author{U. Bovensiepen}\email[] {uwe.bovensiepen@uni-due.de}
\affiliation{Faculty of Physics and Center for
Nanointegration (CENIDE), University of Duisburg-Essen,
Lotharstr.~1, 47057 Duisburg, Germany}

\date{\today}

\begin{abstract}
The transport of optically excited, hot electrons in heterostructures is analyzed by femtosecond, time-resolved two-photon photoelectron emission spectroscopy (2PPE) for epitaxial Au/Fe/MgO(001). We compare the temporal evolution of the 2PPE intensity upon optically pumping Fe or Au, while the probing occurs on the Au surface. In the case of Fe-side pumping, assuming independent relaxation in the Fe and Au layers, we determine the hot electron relaxation times in these individual layers by an analysis of the Au layer thickness dependence of the observed, effective electron lifetimes in the heterostructure. We show in addition that such a systematic analysis fails for the case of Au-side pumping due to the spatially distributed optical excitation density, which varies with the Au layer thickness. This work extends a previous study [Beyazit et al., Phys. Rev. Lett. \textbf{125}, 076803 (2020)] by new data leading to reduced error bars in the determined lifetimes and by a non-linear term in the Au-thickness dependent data analysis which contributes for similar Fe and Au film thicknesses.
\end{abstract}

\maketitle

\section{Introduction}

Excited electrons in Bloch bands of condensed matter scatter on femto- to picosecond timescales due to the strong interaction with bosons and other electrons by e-boson and e-e scattering, respectively \cite{Brorson1987,Shah99,chulkov_2006,Bauer15}. Due to the filled valence band in semiconductors, e-phonon coupling dominates the relaxation dynamics in the conduction bands of these materials. In metals, the half-filled bands provide a large phase space for e-e scattering. Besides the scattering rates as a function of electron energy $E$ and momentum $\textbf{k}$, the propagation of such excitations in real space is important to consider from a fundamental point of view as well as in device applications. Finally, gradients in real space induce currents of free charges \cite{Brorson1987,Hohlfeld2000} which will heat up the crystal lattice and the device structure by e-ph coupling.

While the decay rates of electronic excitations in metals are by now rather well understood \cite{chulkov_2006,Bauer15}, the nature of currents on femto- and picosecond timescales are a topic of current research and fundamental questions are of interest. In a seminal study using femtosecond laser pulses in a pump-probe experiment on freestanding Au films, Brorson et al. have concluded on the ballistic nature of electron currents \cite{Brorson1987}. On the other hand, it was shown more recently that the analysis of propagation velocities calculated by dividing the film thickness by the propagation time is not accounting for the actual electronic propagation pathway \cite{liu_2005,beyazit_2020,melnikov_2022} and individual e-e scattering processes occur although the determined velocities along the film's normal direction are close to the Fermi velocity \cite{beyazit_2020}. Such transport phenomena can have considerable influence in the quantitative analysis of hot electron lifetimes \cite{aeschlimann_APA00,liso_APA04b} and transient electron distribution functions \cite{liso_APA04} if discarded in surface sensitive methods like photoelectron emission spectroscopy.

Ultrafast electron currents in metals can carry a spin polarization and the resulting spin currents and their interaction with ferromagnetic layers in heterostructures have provided new opportunities to control magnetic excitations mediated by spin-pumping, spin-accumulation, and spin-transfer torque \cite{malinowski_2008, Melnikov2011, bergeard_16, Razdolski2017a, Hellmann_RevModPhys_2017, nenno_18,chen_2019, ortiz_2022}. These studies rely on magneto-optical or THz probes which do not directly access the transient electron distributions. To complement those studies it is therefore desired to provide energy- and time-resolved information on the propagating electrons. The specific time- and energy-dependent electron distributions will provide input to works that so far assumed thermalized distributions \cite{beens_2022} or lack sensitivity to non-thermal electrons \cite{buehlmann_2020}.

In this article we report time-resolved two-photon photoelectron emission spectroscopy (2PPE) results on epitaxial Au/Fe/MgO(001) heterostructures. We compare directly the 2PPE spectra detected on the Au surface in case of Fe- and Au-side pumping and analyze the energy-dependent relaxation and propagation times as a function of the Au layer thickness $d_{\mathrm{Au}}$. In case of Fe-side pumping the determined relaxation times depend on the energy above the Fermi level $E-E_{\mathrm{F}}$ and on $d_{\mathrm{Au}}$, which is qualitatively explained by a sum of the decay rates in Fe and Au following Matthiesen's rule. In case of Au-side pumping this separation fails which is explained by the difference in the optical excitation profiles for Fe- and Au-side pumping. We complement our previous work published in Ref.~\cite{beyazit_2020} by recently obtained results that are reported here in combination with an extension of the fitting model.

\section{Experimental Details}

\subsection{Sample Preparation and Characterization}

Epitaxial Au-Fe heterostructures were grown on MgO(001) by molecular beam epitaxy. Fe(001) was prepared on the MgO(001) substrate followed by Au(001). As described in Refs.~\cite{muehge_1994,Melnikov2011,mattern_2022} the in-plane axes of both layers are rotated by $\pi/4$ with respect to each other to minimize the lattice mismatch between Fe and Au which facilitates pseudomorphic growth with an atomically sharp interface. The MgO(001) substrates of $10\times10$ mm$^2$ were cleaned in ultrasonic baths of ethanol, isopropanol, and acetone. Subsequently, they were put into ultrahigh vacuum and exposed to O$_2$ at a partial pressure of $2\cdot10^{-3}$~mbar at a temperature of 540~K to remove carbon contamination. The Fe layer and the first nanometer of Au were evaporated at 460~K. Then, the sample was cooled to room temperature and the remaining Au was evaporated in the following step-wedge structure. We varied $d_{\mathrm{Au}}$ systematically from 5 to 70~nm by growing a wedge shaped Au layer with 17 steps on a 7~nm thick Fe layer. Each step is 0.4~mm wide and can be accessed specifically by the laser pulses focused to a spot of $140\pm30\mu$m diameter FWHM for the visible pump and $90\pm30\mu$m FWHM for the ultraviolet probe. The thickness of the Au and Fe layers were determined by a TOF-SIMS analysis. The error in the film thickness determination is $\pm 10$\% for all films but for $d_{\mathrm{Au}}=7$~nm, where it is $\pm 20$\%.

\subsection{Time-resolved Two-photon Photoelectron Emission Spectroscopy}

\begin{figure}[t]
    \centering
    \includegraphics[width=0.99\columnwidth]{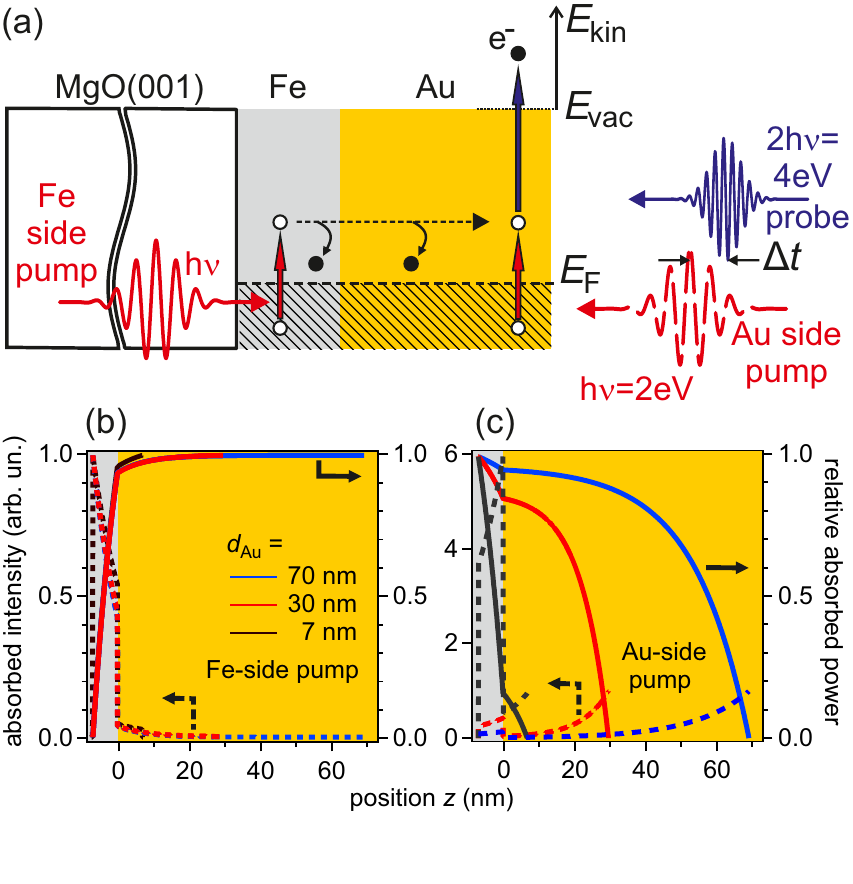}
    \caption{(a) Experimental geometry of the pump-probe experiment with Fe- and Au-side pumping of Au/Fe/MgO(001) heterostructures by visible femtosecond laser pulses at $h\nu=2.1$~eV photon energy. Probing occurs by analysis of the two-photon photoelectron emission spectrum at the Au surface induced by $2h\nu=4.2$~eV. In case of Fe-side pumping electrons have to propagate to the Au surface before being probed. Panels (b,c) illustrate the spatial distribution of the excitation density by the calculated relative absorbed pump light intensity with respect to the incident light field (dashed lines, left axis) and the relative absorbed pump light intensity (solid lines, right axis) as a function of Au layer thickness for (b) Fe- and (c) Au-side pumping. The thickness of the Fe layer is 7 nm (gray area). Note that the nominal maximum absorption intensity in (c) is higher at the interface than in (b) because of the change in optical constants for a change in layer sequence.}
    \label{fig:fig1}
\end{figure}

Femtosecond laser pulses are generated by a commercial regenerative Ti:sapphire amplifier (Coherent RegA 9040) combined with a non-collinear optical parametric amplifier (NOPA, Clark-MXR) operating at 250~kHz repetition rate tuned to a signal photon energy of $h\nu=$2.1~eV, which we frequency doubled in a BBO crystal subsequently. We use these pairs of femtosecond (fs) pulses at 2.1 and 4.2~eV with pulse durations below 40~fs at a time delay $\Delta t$ as pump and probe pulses, respectively. The probe pulses reach the Au surface at 45 deg. angle of incidence. Since the transparent MgO(001) substrate allows direct optical excitation of the Fe layer, the pump pulse can be sent to either the Au or the Fe side of the sample by different pathways, also at 45 deg. angle of incidence. Thereby, pump and probe laser pulses propagate within their foci simultaneously over the sample which avoids a deterioration of time resolution. Typical incident fluences are $50~\mu$J/cm$^2$ and $1~\mu$J/cm$^2$ for pump and probe, respectively. The sample was kept in ultrahigh vacuum and at room temperature. Photoelectrons are detected in normal emission direction from the Au surface by a self-built electron time-of-flight spectrometer \cite{kirchmann_APA08}. The concept of the 2PPE experiment with a direct comparison of Fe-side and Au-side pumping of the Au/Fe/MgO(001) heterostructure is illustrated in Fig.~\ref{fig:fig1}(a).

Optical absorption of the pump pulses in the heterostructure differs for Fe- and Au-side pumping, since the absorption coefficient of 2~eV photons in Fe is considerably larger than in Au. We have calculated the absorption of the pump pulse in Au/Fe by the electric field inside the material in both pump configurations using the IMD software \cite{imd_1998} and derive the absorbed power $P(z)$ in the different constituents with $z$ being the interface normal direction; $z=0$ is set to the Fe-Au interface.

\begin{equation}
P(z) = n(z)I(z) = n(z)|E(z)|^2 = n(z)I_{0}e^{-\alpha z},
\label{eq:1}
\end{equation}

here $\alpha$ is the absorption coefficient, $I(z)$ the intensity, $E(z)$ the electric field, and $n(z)$ the refractive index in the respective material. The power in the layer stack is given by the real part of the Poynting vector. Fig.~\ref{fig:fig1}(b,c) depict the relative absorbed intensities for (b) Fe-side and (c) Au-side pumping (dashed lines, left axes). The relative absorbed power was obtained by subtracting the transmitted field and normalizing the incoming intensity to unity because we focus on the attenuation of the field by absorption (solid lines, right axes). We considered the pump photon energy of 2.1~eV, the angle of incidence $\theta = 45^{\circ}$, the $p$-polarization of the light, the refractive indices $n_{\mathrm{Au}}$(2.1~eV) = 0.25, $n_{\mathrm{Fe}}$(2.1~eV) = 2.91 and the extinction coefficients $k_{\mathrm{Au}}$(2.1~eV) = 3.07, $k_{\mathrm{Fe}}$(2.1~eV) = 3.02~\cite{palik_1998} with $\alpha=4\pi k/\lambda$; $\lambda$ is the optical wavelength. We find that in case of Fe-side pumping the absorption is dominated by the Fe layer: 96\% for $d_{\mathrm{Au}} = 7$~nm and 94\% for $d_{\mathrm{Au}} = 70$~nm. In the case of Au-side pumping the situation is diverse. A pronounced variation in the relative intensity occurs at the Fe-Au interface, where the refractive index changes and the light field is attenuated in Au before it reaches Fe. Note, that for $d_{\mathrm{Au}} = 7$~nm the dominant absorption occurs in Fe though it reaches Au first. For thicker Au layers the pump absorption is distributed over the Au layer.

\begin{figure}[t]
    \centering
    \includegraphics[width=0.99\columnwidth]{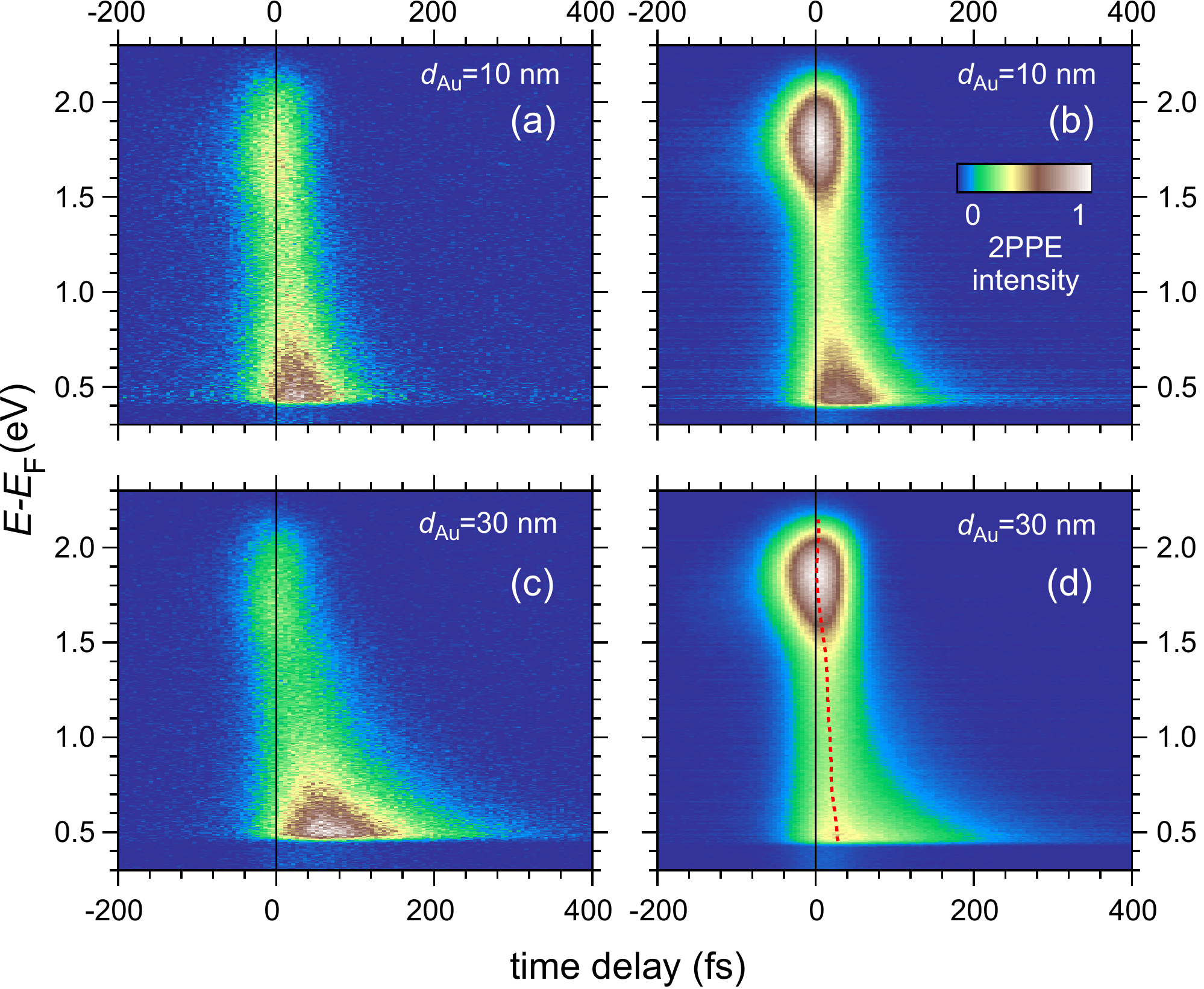}
    \caption{Representative 2PPE intensity as a function of time delay and energy above the Fermi level in a false color representation for $d_{\mathrm{Fe}}=7$~nm and $d_{\mathrm{Au}}=10$ and 30~nm, as indicated; (a,c) Fe-side pumping, (b,d) Au-side pumping. The red, dashed line in panel (d) indicates the 2PPE intensity maxima for different energies with time delay.}
    \label{fig:fig2}
\end{figure}

\section{Experimental Results and Data Analysis}

The time-resolved 2PPE intensity was measured as a function of $d_{\mathrm{Au}}$ on the step-wedged sample and representative results are shown in Fig.~\ref{fig:fig2} after subtraction of time-independent contributions originating from multi-photon photoemission within a single pump or probe pulse, which is typically 10\% of the time-dependent intensity. Characteristic changes with increasing  $d_{\mathrm{Au}}$ for Fe-side pumping are a reduced intensity at $E-E_{\mathrm{F}}=1.7$~eV and a shift of the intensity maximum at lower energy near $E-E_{\mathrm{F}}=0.6$~eV to later $\Delta t$. For Au-side pumping the maximum intensity is found close to the top end of the spectrum at 1.7~eV which exhibits a decay to nominally negative time delays, which represents an excitation sequence of pumping with 4.2~eV and probing with 2.1~eV. The energy of the spectral peak corresponds therefore to $E-E_{\mathrm{F}}=3.8$~eV and is assigned to a manifold of image potential states. The individual image potential states are not resolved, likely due to inhomogeneous broadening and the remaining surface inhomogeneity. The temporal evolution of the spectral contributions do not change with $d_{\mathrm{Au}}$ besides intensity variations.

For all data sets an increase in intensity towards lower $E-E_{\mathrm{F}}$ and later $\Delta t$ is recognized, which can be weaker or stronger depending on $d_{\mathrm{Au}}$ or the pumping geometry. This effect has two origins. (i) The hot electron lifetime increases according to Fermi-liquid theory $\propto \left(E-E_{\mathrm{F}}\right)^{-2}$ \cite{chulkov_2006,Bauer15}. (ii) At electron energies $E-E_{\mathrm{F}}$ below half of the pump photon energy, secondary electrons will contribute to the 2PPE intensity \cite{liso_APA04b}. In this work we focus on the analysis of the hot electron lifetimes, i.e. the inverse rate of the primary inelastic scattering event, for which 2PPE is the appropriate method. The contribution of secondary electrons and effects towards electron thermalization upon Fe- and Au-side pumping was reported recently in Ref.~\cite{kuehne_2022} based on time-resolved linear photoelectron emission spectroscopy.

For both pump geometries $\Delta t = 0$ was determined for consistency reasons by the 2PPE intensity maximum at the highest energy at the top end of the spectrum. We note that there is a certain ambiguity in this choice of time zero. For Fe-side pumping a certain propagation time of the excited electron through the layer stack before the electron is detected at the Au surface occurs. For Au-side pumping, the spectrally broad image potential state feature could extend up to the top end of the spectrum. As a consequence, the chosen $\Delta t = 0$ would be shifted to negative time delays since the finite decay time of the image potential states convoluted with the pulse duration results in an effective shift of the intensity maximum since the trailing part of the probe laser pulse contributes to the signal. To estimate this potential inaccuracy in the determination of $\Delta t = 0$, we indicate in Fig.~\ref{fig:fig2}(d) the intensity maxima as a function of $\Delta t$ for all energies by a dashed red line. At $E-E_{\mathrm{F}}=1.5$~eV to 1.0~eV the maximum is shifted to +15~fs without a clear decay. We consider in the following that the actual time zero is uncertain within this interval of 0 to 15~fs.

\begin{figure}[t]
    \centering
    \includegraphics[width=0.99\columnwidth]{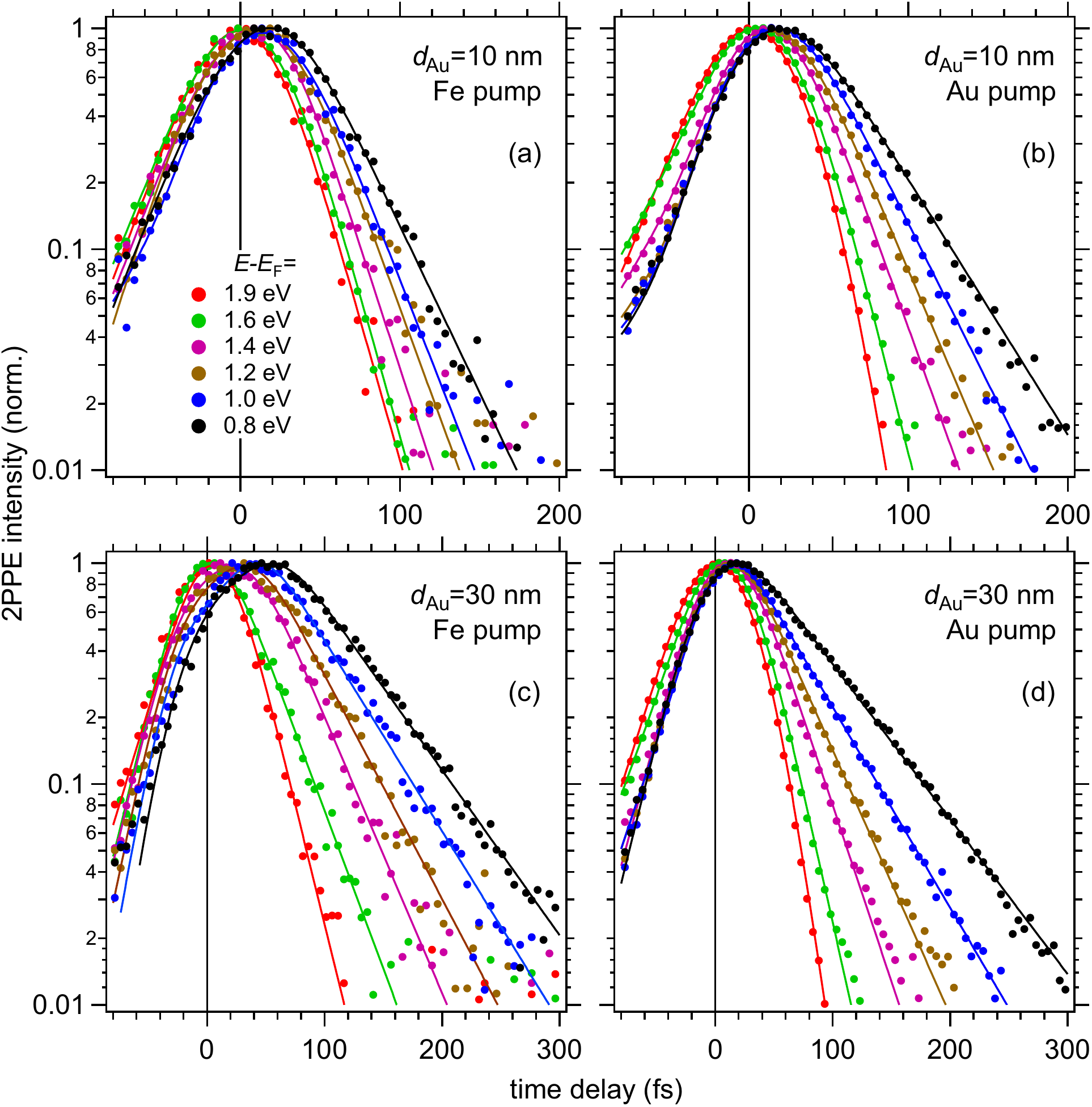}
    \caption{Normalized 2PPE intensity as a function of time delay at different energies $E-E_{\mathrm{F}}$ for $d_{\mathrm{Fe}}=7$~nm and $d_{\mathrm{Au}}$ and pumping as indicated in panels (a-d). The dots are experimental data, the lines represent least-square fits using an exponential relaxation convoluted with a Gaussian, see text for details. Note the different intervals in time delay are depicted in panels (a,b) and (c,d).}
    \label{fig:fig3}
\end{figure}

Similar measurements were taken for 13 different $d_{\mathrm{Au}}$ in case of Fe-side pumping and 7 different $d_{\mathrm{Au}}$ in case of Au-side pumping. We analyzed the time-dependent 2PPE intensities at constant energy $E-E_{\mathrm{F}}$ for all these measurements after normalization to the time-dependent peak maximum. Typical examples of such datasets are shown in Fig.~\ref{fig:fig3} for the data of Fig.~\ref{fig:fig2}. A comparison of these traces for $d_{\mathrm{Au}}$ in panel (c,d) for both pumping geometries highlights that the time delays of the intensity maxima are shifted for Fe-side pumping towards later $\Delta t$ much more than for Au-side pumping. This effect is attributed to propagation of the electronic excitation through the Au layer and quantified by a time offset $t_0$, see Ref.~\cite{beyazit_2020} for a discussion of the results on $t_0$. These traces are fitted for both pumping geometries with two exponential decays, one towards negative $\Delta t$ to account for the contribution excited by 4.2~eV photons, and one for the decay towards positive $\Delta t$ excited by 2.1~eV photons shifted by $t_0$. All is convoluted with a Gaussian of $\sim 50$~fs width to account for the cross-correlation of the laser pulses. The resulting fits are included in Fig.~\ref{fig:fig3} by solid lines.

For a discussion of the determined lifetimes it is important to consider that the excited electron can relax in the Fe or in the Au layer of the heterostructure. If it does not relax in Fe we assume in the following analysis that the electron is injected elastically into the Au layer where it relaxes subsequently. This is rationalized with the single crystalline interface structure \cite{Melnikov2011} and by the fact that no interface decay contribution is required to describe the data, as discussed below. For the Fe-side pumping and Au-side probing geometry the detected electron has to propagate through the layer stack and the probability for an electron to decay will increase in both layers with their respective thicknesses. In the limit of a sufficiently thin Fe layer, which is grown on the insulating MgO(001), transport effects can be neglected in Fe. Supposing that the Au layer grown on top of Fe is sufficiently thin that its contribution to the decay probability can be neglected, this experimental geometry would be expected to measure the relaxation dynamics in Fe. Vice versa, if the Fe layer is very thin and the Au layer is sufficiently thick, the relaxation dynamics in Au would be measured. As a function of $d_{\mathrm{Au}}$ we can therefore expect to detect a combination of relaxation in the Fe and Au constituents as suggested by Mathiessens's rule. The data presented in Fig.~\ref{fig:fig3} show clearly a faster relaxation for $d_{\mathrm{Au}}=10$ nm compared to $d_{\mathrm{Au}}=30$ nm for both, Fe-side and Au-side pumping. In order to test this hypothesis we analyze the inverse relaxation times as a function of $d_{\mathrm{Au}}$ as a linear function of $1/d_{\mathrm{Au}}$ to account for the variation of the decay contribution in Au by

\begin{equation}
\frac{1}{\tau^{\mathrm{eff}}(d_{\mathrm{Au}})}=A+\frac{B}{d_{\mathrm{Au}}}.
\label{eq:2}
\end{equation}

\begin{figure}[t]
    \centering
    \includegraphics[width=0.99\columnwidth]{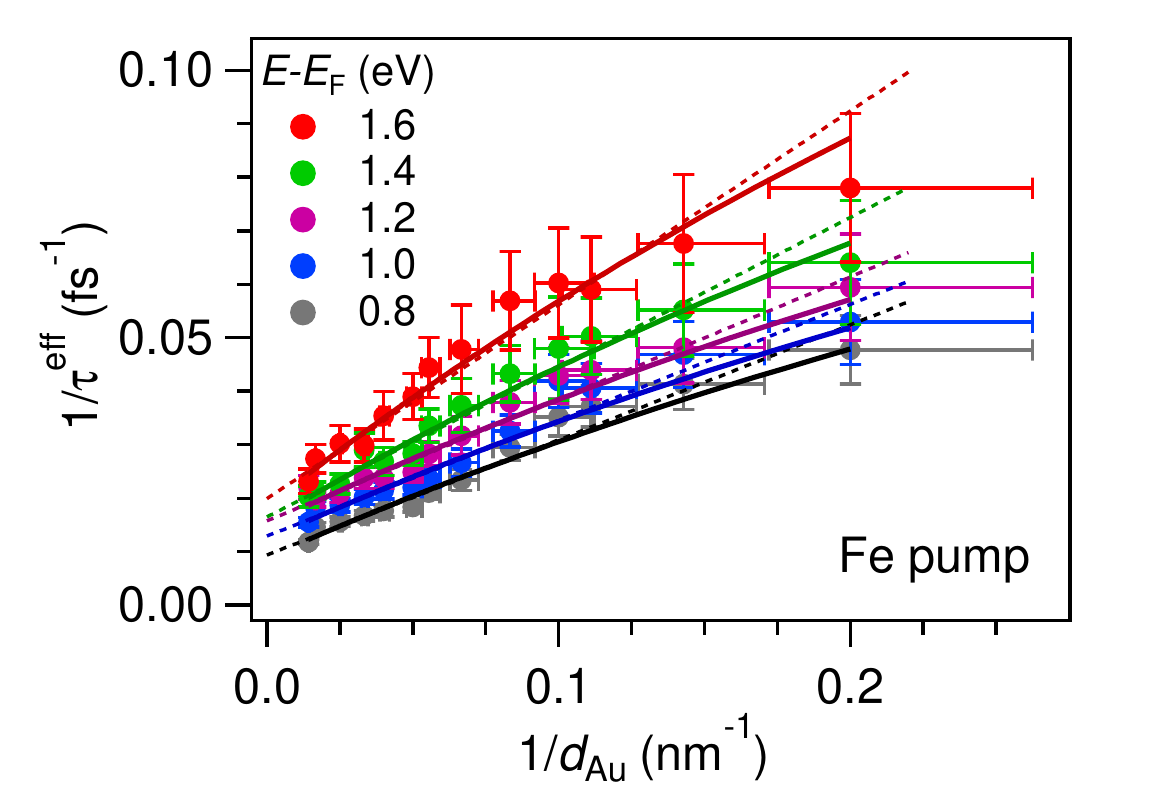}
    \caption{The relaxation rates determined from the inverse hot electron lifetimes in case of Fe-side pumping are plotted as filled circles for energies above $E_{\mathrm{F}}$ as a function of the inverse Au layer thickness. The solid (dotted) lines depict non-linear (linear) functions fitted to the relaxation rates at the energy represented by the indicated color code. See the text for details.}
    \label{fig:fig4}
\end{figure}

In this empirical limit, $1/A=\tau_{\mathrm{Au}}$ and $1/B=\tau_{\mathrm{Fe}} / d_{\mathrm{Au}}^0$, where $d_{\mathrm{Au}}^0$ normalizes $d_{\mathrm{Au}}$ and is chosen as 1~nm. Fig.~\ref{fig:fig4} shows the respective analysis. The linear functions of Eq.~\ref{eq:2} are included as dotted lines which fit the experimental data reasonably well within the error bars. At large $1/d_{\mathrm{Au}}$ we observe systematic deviations of the linear fit from the experimental. The fit tends to overestimate the measured relaxation rates. In the following we derive a suitable non-linear correction. The corresponding fits are plotted as solid lines in Fig.~\ref{fig:fig4}.

Under the approximation that the electronic velocity in Fe and Au is similar and that scattering at the Fe-Au interface can be discarded, the scattering probability in the layer stack $1/\tau^{\mathrm{eff}}$ is determined by $d_{\mathrm{Au}}$ and $d_{\mathrm{Fe}}$ \cite{beyazit_2020} following,

\begin{equation}
\frac{d_{\mathrm{Au}}+d_{\mathrm{Fe}}}{\tau^{\mathrm{eff}}}=\frac{d_{\mathrm{Au}}}{\tau_{\mathrm{Au}}}+\frac{d_{\mathrm{Fe}}}{\tau_{\mathrm{Fe}}}.
\label{eq:3}
\end{equation}

\noindent Since $d_{\mathrm{Au}}$ was varied and $d_{\mathrm{Fe}}$ was kept constant at 7~nm

\begin{equation}
\frac{1}{\tau^{\mathrm{eff}}(d_{\mathrm{Au}})}=\frac{d_{\mathrm{Au}}}{d^{\mathrm{eff}}_{\mathrm{Fe}}+d_{\mathrm{Au}}} \cdot \frac{1}{\tau_{\mathrm{Au}}}+\frac{d^{\mathrm{eff}}_{\mathrm{Fe}}}{d^{\mathrm{eff}}_{\mathrm{Fe}}+d_{\mathrm{Au}}} \cdot \frac{1}{\tau_{\mathrm{Fe}}}.
\label{eq:4}
\end{equation}

\noindent As discussed below in Sec.~IV we introduce $d^{\mathrm{eff}}_{\mathrm{Fe}}$ to account for the optical inhomogeneous pumping of Fe and for different electron velocities in Fe and Au. For $d_{\mathrm{Au}} \gg d^{\mathrm{eff}}_{\mathrm{Fe}}$ the factor of the first term becomes one and the one of the second term tends to reduce $1/d_{\mathrm{Au}}$ as in Eq.~\ref{eq:2}. For $d_{\mathrm{Au}} \approx d^{\mathrm{eff}}_{\mathrm{Fe}}$ the fitting by Eq.~\ref{eq:4} overcomes the previous deviation between experimental data and the fitting using Eq.~\ref{eq:2} for thin Au films, see Fig.~\ref{fig:fig4}.

\begin{figure}[t]
    \centering
    \includegraphics[width=0.99\columnwidth]{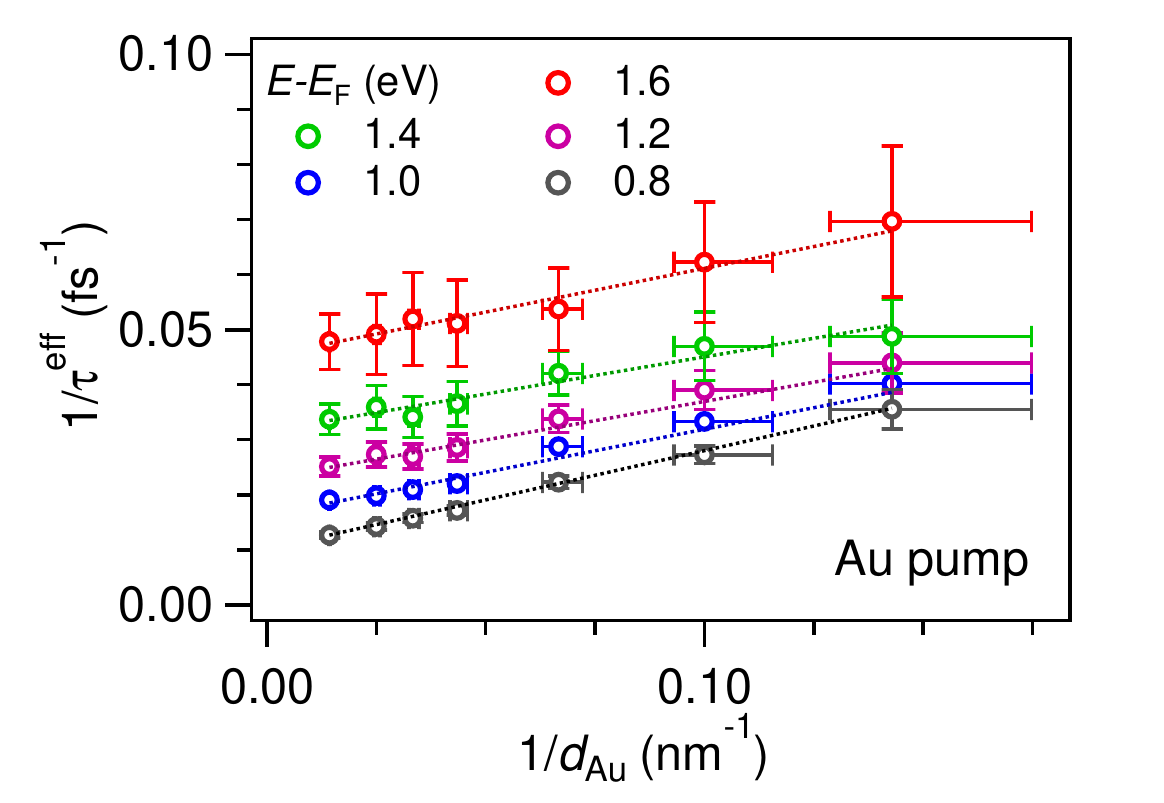}
    \caption{The relaxation rates determined from the inverse hot electron lifetimes in case of Au-side pumping, see text for details, are plotted as open circles for energies above $E_{\mathrm{F}}$ as a function of the inverse Au layer thickness. The dotted lines are linear fits to the relaxation rates at the energy represented by the indicated color code.}
    \label{fig:fig5}
\end{figure}

This analysis potentially also holds for the experiments which employ Au-side pumping. To test this hypothesis, we plot $1/\tau^{\mathrm{eff}}$ obtained for Au-side pumping as a function of $1/d_{\mathrm{Au}}$ and fit the data by Eq.~\ref{eq:2}. The results are shown in Fig.~\ref{fig:fig5}. In this case the linear fitting describes the data very well, but the quantitative behavior of the results for $A$ and $B$ as a function of $E-E_{\mathrm{F}}$ differs from the results obtained for Fe-side pumping, c.f. Fig.~\ref{fig:fig4}. For Au-side pumping the slope $B$ increases with decreasing energy while the opposite trend is found for Fe-side pumping. In addition, the offset $A$ is about two times larger than for Fe-side pumping.

\begin{figure}[t]
    \centering
    \includegraphics[width=0.99\columnwidth]{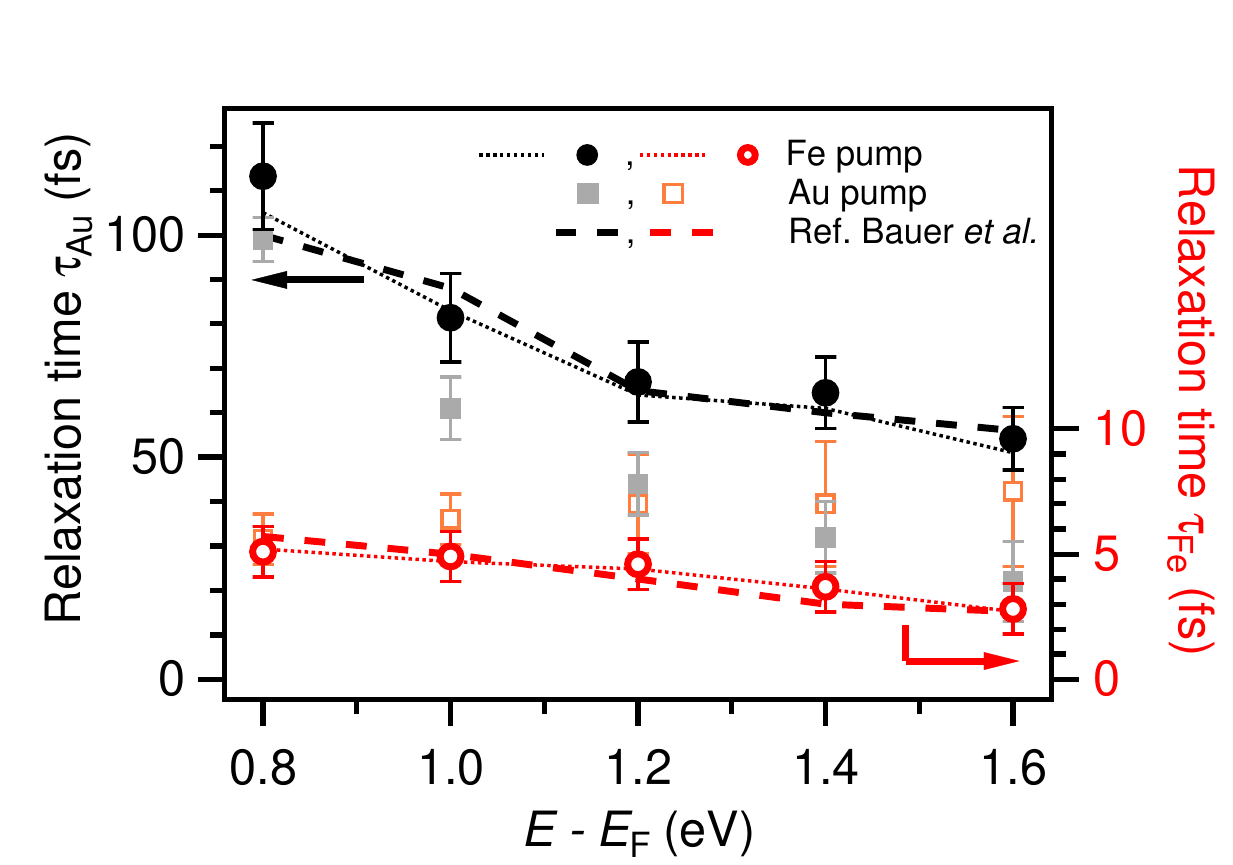}
    \caption{Filled and open circles are the best fit results of $\tau_{\mathrm{Fe}}$ and $\tau_{\mathrm{Au}}$ by the Au-thickness dependent analysis of the hot electron lifetimes in case of Fe-side pumping using the non-linear fitting analysis. The results for the linear fitting are depicted by thin dotted lines. Open and filled squares are the corresponding results for Au-side pumping. Dashed lines represent literature data for hot electron lifetimes for bulk Fe and Au \cite{Bauer15}. Black and gray data for $\tau_{\mathrm{Au}}$ are referred to the left axis, red and orange data for $\tau_{\mathrm{Fe}}$ to the right axis.}
    \label{fig:fig6}
\end{figure}

The results for $\tau_{\mathrm{Au}}$ and $\tau_{\mathrm{Fe}}$ obtained for the linear and non-linear analysis in the Fe- and for the linear analysis in the Au-side pumping geometry are compiled as a function of energy in Fig.~\ref{fig:fig6}. We also include literature data for hot electron lifetimes in bulk Au and Fe taken from Bauer et al.~\cite{Bauer15} for comparison. The findings for $\tau_{\mathrm{Au}}$ and $\tau_{\mathrm{Fe}}$ in case of Fe-side pumping is in very good agreement with the literature data, which holds for both fitting models. We conclude that for Fe-side pumping a separation of the relaxation dynamics in the Fe and Au constituents is successful. We note that the non-linear fitting required to assume a value of $d^{\mathrm{eff}}_{\mathrm{Fe}}=1.3\pm0.2$~nm which is much smaller than the actual Fe film thickness of $d_{\mathrm{Fe}}=7$~nm to obtain this agreement. This result is discussed in the Sec.~IV below. In case of Au-side pumping the deviation of the experimental data points and the literature data is significant for $E-E_{\mathrm{F}}>0.8$~eV. In this experimental geometry the separation of the dynamics in the two constituents fails.

\section{Discussion}

Hot electron transport phenomena have been widely identified and discussed in the literature of pump-probe experiments in which pump and probe pulses impinge at the sample surface from the identical side, see, e.g., \cite{aeschlimann_APA00,Hohlfeld2000,liso_APA04,liso_APA04b,malinowski_2008,schellekens_2014,wieczorek_2015,chen_2017,chen_2019}. We showed in this work that profound differences occur in the quantitative analysis for two experimental configurations in which pump and probe pulses arrive at the same or opposite sides of the sample surface. These differences are not primarily related to the analysis of the transport phenomenon itself but have effects on the determined hot electron lifetimes. This aspect might be particularly relevant in heterostructure samples where electrons can be transferred among different constituents. We explain these differences between Fe- and Au-side pumping by the different pump absorption profile in the heterostructure. As shown in Fig.~\ref{fig:fig1}(b,c) the absorption profile is very different for the two pumping geometries. For the Fe-side pumping, the hot electrons are primarily excited in Fe and propagate a well-defined distance to the Au surface, where the photoelectrons are detected. In the Au-side pumping situation, the electrons are excited in Fe and Au with different spatial profiles depending on $d_{\mathrm{Au}}$. Since the electronic transport processes are determined by gradients in excitation density and electronic temperature \cite{Hohlfeld2000}, the velocity distributions and the corresponding transient electron densities will differ for the two pumping geometries. Therefore, it is not only the opposite direction of the pump-induced transport that distinguishes Au and Fe-side pumping. The quantitative difference in the absolute values of the spatial gradients of the excitation density leads to spatial redistribution of transient electron density which were already earlier recognized to result in misleading lifetime analysis \cite{aeschlimann_APA00,liso_APA04b}. The wider spatial distribution of the excited electrons that cover Fe and Au in case of Au-side pumping inhibits the separation of the dynamics in this case. So far, the back-side pump and front-side probe configuration proved to be a very suitable approach to analyze electron scattering and transport for the material system under study.

The non-linear fitting approach in the analysis of the hot electron lifetimes for Fe-side pumping resulted in the finding, that $d^{\mathrm{eff}}_{\mathrm{Fe}}=1.3$~nm much thinner than the actual $d_{\mathrm{Fe}}=7$~nm. Here, we discuss two aspects in this context to rationalize this result. (i) Fig.~\ref{fig:fig1}(b) depicts the variation of absorbed pump intensity which changes two times across the Fe layer. A corresponding spatial distribution of excited  electrons are injected from Fe to Au. In consequence, the average distance an electron has to travel before it reaches the Fe-Au interface is 3--4~nm. (ii) In deriving Eq.~\ref{eq:4} we assumed that the electronic velocities in Fe and Au are similar. In fact, they differ at the relevant energies considerably according to GW calculations \cite{Zhukov2006}. For majority electrons in Fe the velocity is 0.6 of the one in Au at $E-E_{\mathrm{F}}=1.5$~eV. To compensate for this difference in velocity a correspondingly thinner Fe sheet might be considered as effective since in the time interval before scattering occurs an electron can cover at a lower velocity a shorter pathway. We discard here minority electrons in Fe since the injection probability across the Fe-Au interface favors injection of majority electrons \cite{Alekhin2017}. Both these effects reduce the effective Fe film thickness in hot electron  injection to Au and the result obtained for $d^{\mathrm{eff}}_{\mathrm{Au}}$ is very plausible.

Finally, we discuss potential impact of our work on other problems. Our work quantifies the energy dependent electron dynamics upon Au- and Fe-side optical excitation. It complements a recently published study \cite{kuehne_2022} which used linear time-resolved photoelectron detection to study the electron distribution near $E_{\mathrm{F}}$. Both these works might serve as input for electron dynamics in the description of optically excited spin currents in metallic heterostructures. A description of these currents by thermal models can be considered as simplified, though they potentially describe the spin currents reasonably well.

An extension of the presented approach to spin-resolved photoelectron spectroscopy appears as promising given the recent development of efficient spin-resolved photoelectron analysis \cite{schoenhense_2015}. A 2PPE experiment will very likely be more suitable than detection in time-resolved linear photoelectron emission \cite{buehlmann_2020}, since it provides higher countrates of photoelectrons that carry time-resolved information.

Previous interest in time-resolved spectroscopic information in layered perovskite systems upon optical excitation at the opposite sample side \cite{Sung2020} suggests an extension of the approach reported in the present work to further material systems. Our study can be considered to address a rather favorable problem since due to the excellent interface structure the scattering centers at the interface were sufficiently small in density that they did not have to be taken into account explicitly. For other material systems this might be different, in particular for structures that promise technological relevance like, e.g., the mentioned perovskite systems. In such a situation the variation of two film thicknesses in the heterostructure promises to analyze the interface scattering specifically \cite{beyazit_2020}. Another potential direction of experiments that build on our demonstration is the use of Au/Fe/MgO(001) as electrodes for hot electrons injected into solid layers prepared on top of Au. This is a promising future research opportunity for solid layered materials. It is also interesting to consider this approach in combination with liquid electrolytes. In this case the detection of photoelectrons could be very challenging, but the detection by a surface sensitive non-linear optical technique might be viable \cite{Melnikov2011}.

\section{Conclusions}

The ultrafast transport of optically excited hot electrons in epitaxial Au/Fe/MgO(001) heterostructures is studied by systematically varying the Au layer thickness. The results presented here extend our earlier study on this heterostructure and enabled a determination of hot electron lifetimes for the Fe and Au constituents separately with reduced error. This analysis implies a non-linear dependence of the effective relaxation rate on the Au layer thickness. It was successful in attributing a systematic deviation at smaller Au layer thickness, in the range of the Fe layer thickness, from its dependence on thicker Au layers for the Fe-side pumping configuration. The success of the analysis in case of Fe-side pumping is attributed to its ability to excite carriers in a specific part in the heterostructure independently on the propagation pathway before detection. Front- and back-side pumping studies are commonly used for studying transport induced effects in nanostructures, however, we showed here that the spatial distribution of the excitation density plays a strong role in modifying the measured results. The results presented here will serve as input for electron dynamics in the description of spin currents in metallic heterostructures, their interfaces, and hot-electrons injected into other coupled systems which are of technological importance.

\begin{acknowledgments}

Funding by the Deutsche Forschungsgemeinschaft (DFG, German Research Foundation) through Project No. 278162697 - SFB 1242, through Project No. BO1823/12 - FOR 5249 (QUAST), and under Germany's Excellence Strategy - EXC 2033 - 390677874 - RESOLV is gratefully acknowledged. We are also grateful for fruitful discussions with J. Beckord.

\end{acknowledgments}

\providecommand{\noopsort}[1]{}\providecommand{\singleletter}[1]{#1}%

\end{document}